\documentclass[11pt,twocolumn]{article}
\usepackage{latexsym,psfig,epsfig,graphicx,subfigure,doublespace}
\makeatletter
\newcommand{\tinyspacing}{\let\CS=\@currsize\renewcommand
{\baselinestretch}{.9}\tiny\CS}
\newcommand{\singlespacing}{\let\CS=\@currsize\renewcommand
{\baselinestretch}{1}\tiny\CS}
\newcommand{\thisspacing}{\let\CS=\@currsize\renewcommand
{\baselinestretch}{1.15}\tiny\CS}
\newcommand{\doublespacing}{\let\CS=\@currsize\renewcommand
{\baselinestretch}{1.2}\tiny\CS}
\newcommand{\largespacing}{\let\CS=\@currsize\renewcommand
{\baselinestretch}{1.3}\tiny\CS}
\makeatother

\newfont{\mmk}{cmr6} 
\newfont{\sfbTen}{cmssbx10} 

\def\blackbox{\hfill {\vrule height3pt width4pt depth2pt}}

\newcommand{\eat}[1]{}

\makeatletter
\def\shukyfootnotetext{
          \xdef\@thefnmark{\null}\@footnotetext}
\makeatother

\newcommand{\delte}
{\mbox{$\mathrel{\hbox{$\bigtriangleup$\raise1.5pt\hbox{\hskip-6.5pt
{\tiny $e$}}}}$\,}}
\newcommand{\deltu}
{\mbox{$\mathrel{\hbox{$\bigtriangleup$\raise1.5pt\hbox{\hskip-6.5pt
{\tiny $\mu$}}}}$\,}}

\newcommand{\dele}
{\mbox{$\mathrel{\hbox{$\bigtriangledown$\raise2pt\hbox{\hskip-6.5pt
{\tiny $e$}}}}$\,}}
\newcommand{\delu}
{\mbox{$\mathrel{\hbox{$\bigtriangledown$\raise2pt\hbox{\hskip-6.5pt
{\tiny $\mu$}}}}$\,}}

\newcommand{\monus}{\mbox{$\mathrel{\hbox{$-$\raise2pt\hbox{\hskip 
-5.5pt$\cdot$}}}$\,}}

\newcommand{\semijoin}{\mbox{$\mathrel{\raise1pt\hbox{\vrule height5pt
depth0pt\hskip-1.5pt$>$\hskip -2.5pt$<$}}$}}

\newtheorem{alg}{Algorithm}[section]
\newtheorem{conj}{Conjecture}[section]
\newtheorem{corol}{Corollary}[section]
\newtheorem{defin}{Definition}[section]
\newtheorem{ex}{EXAMPLE}[section]
\newtheorem{lemm}{Lemma}[section]
\newtheorem{ob}{Observation}

\newtheorem{prop}{Proposition}[section]
\newtheorem{thm}{Theorem}[section]

\newcommand{\bdisplay}{\begin{verse}\begin{tabbing}}
\newcommand{\edisplay}{\end{tabbing}\end{verse}}

\newenvironment{definition}[1]{\begin{defin}\begin{rm}({\bf #1})}{{\hfill$\Box$}\end{rm}\end{defin}}

\newenvironment{example}{\begin{ex} \nopagebreak
  \begin{rm}}{{\hfill$\Box$}\end{rm}\end{ex}} 

\newenvironment{lemma}{\begin{lemm}}{{\hfill$\Box$}\end{lemm}}

\newenvironment{proposition}{\begin{prop}}{{\hfill$\Box$}\end{prop}}

\newenvironment{theorem}{\begin{thm} \nopagebreak}{{\hfill$\Box$}\end{thm}}

\newcounter{linenumber}[alg]

\newcounter{clauses}
\newcounter{rule}[clauses]

\newcounter{sql} \newcounter{sqlstatement}[sql]

\def \ip#1{\par\penalty-5000\noindent\hangindent20pt\hangafter1
\hbox to 20pt{#1\hfill}\ignorespaces}

\newif\ifproof
\prooffalse

\newif\ifproofs
\proofsfalse

\newcommand{\scom}[1]{}

\newcounter{temppage}

\def\papertype #1 raised #2 {
\vspace{-#2}
\vbox to 0pt{\hfill\framebox{\bf #1}}
\vspace{#2}
}

\newlength{\lineheight}
\renewcommand{\baselinestretch}{1.0}
\newsavebox{\savepar}
{\end{minipage}\end{lrbox}\fbox{\rule{0cm}{1cm}{\usebox{\savepar}}}}

\newcommand{\squishlist}{
  \begin{list}{$\bullet$}
   {
     \setlength{\itemsep}{0pt}
     \setlength{\parsep}{0pt}
     \setlength{\topsep}{0pt}
     \setlength{\partopsep}{0pt}
     \setlength{\leftmargin}{1.5em}
     \setlength{\labelwidth}{1em}
     \setlength{\labelsep}{0.5em} } }
\newcommand{\squishend}{
   \end{list}  }

\begin{document}
\bibliographystyle{alpha}
\title{Database Reformulation with Integrity Constraints \\ (extended abstract)}

\vspace{-1cm}
{\small 
\author{Rada Chirkova\footnote{\noindent This author's work on this material has been supported by the National Science Foundation under Grant No. 0307072}\\
Department of Computer Science\\
North Carolina State University\\
chirkova@csc.ncsu.edu
\and
Michael R. Genesereth \\
Department of Computer Science\\
Stanford University\\ 
genesereth@cs.stanford.edu
}
}

\date{}
\maketitle

\vspace{-1cm}

\begin{abstract}
{\small 
In this paper we study the problem of reducing the evaluation costs of queries on finite databases in presence of integrity constraints, by designing and materializing views. Given a database schema, a set of queries defined on the schema, a set of integrity constraints, and a storage limit, to find a solution to this problem means to find a set of views that satisfies the storage limit, provides equivalent rewritings of the queries {\it under the constraints} (this requirement is weaker than equivalence in the absence of constraints), and reduces the total costs of evaluating the queries. This problem, {\it database reformulation}, is important for many applications, including data warehousing and query optimization. We give complexity results and algorithms for database reformulation in presence of constraints, for conjunctive queries, views, and rewritings and for several types of constraints, including functional and inclusion dependencies. To obtain better complexity results, we introduce an {\it unchase} technique, which reduces the problem of query equivalence under constraints to equivalence in the absence of constraints {\it without} increasing query size.
}
\end{abstract} 

\vspace{-0.5cm}
\section{Introduction}
\vspace{-0.3cm}

\label{intro-section}

In many contexts it is beneficial to answer database queries using derived data called {\it views}. A view is a named query, which can be stored in a database system as a definition ({\it virtual view}) or as an answer to the query ({\it materialized view}). A user query can be answered using views via a new definition that is called a {\it rewriting} and is built in terms of the views. Using virtual or materialized views in query answering~\cite{LevyMSS95} is relevant in applications in information integration, data warehousing, web-site design, and query optimization. Two main directions in answering queries using views are (1) feasibility: to obtain some answer to a given query using given views, as in the information-integration scenario, and (2) efficiency: to reduce query-execution time by using the views, as in the query-optimization scenario. Within the efficiency direction, which is our focus in this paper, the objective is typically to use views to obtain {\it equivalent} query rewritings --- that is, definitions that give the exact answer to the query on all databases. Answering queries using views has been explored in depth for relational database systems~\cite{Kanellakis90} and for conjunctive queries, which can be defined via positive existential conjunctive formulas of first-order logic~\cite{Enderton72}; for a survey of methods for answering queries using views see~\cite{Halevy01}. 

In the past few years, significant research efforts have been concentrated on {\it view selection}, that is, on developing methods for defining and precomputing materialized views to answer predefined queries; existing approaches differ in their main objective (feasibility or efficiency) and in how they explore the search space of views and rewritings for the given queries, typically on finite databases. \cite{ChirkovaG00} introduced the approach of database reformulation, with an emphasis on efficiency and on {\it complete} exploration of the search space of efficient rewritings. Formally, starting with a set of finite database relations and a set of queries, the problem is to design a set of views of the database relations that (1) can be materialized under a given restriction (such as a storage limit, i.e., the amount of disk space available for storing the view relations) and, once materialized, (2) can be used by a given evaluation algorithm in answering the queries equivalently and more efficiently than the original relations. The schema consisting of the materialized views is called a {\it reformulation} of the problem input.  A reformulation is {\it beneficial} (or {\it optimal}) if it is as efficient as or more efficient than the original (or every other) [re]formulation on all given queries and all databases consistent with the given schema. It has been shown~\cite{ChirkovaG00} that there are reformulation problems for which there are infinitely many beneficial reformulations; at the same time, only finitely many of these reformulations need to be considered since any other reformulation is either larger or less efficient to use. Therefore, it is possible to find an optimal reformulation in finite time.

The results in~\cite{ChirkovaG00} do not take into account integrity constraints, or dependencies, on the base relations in the database. Dependencies are semantically meaningful and syntactically restricted sentences of the predicate calculus that must be satisfied by any ``legal" database; examples include functional dependencies and foreign-key constraints~\cite{Kanellakis90,AbiteboulHV95}. The presence of dependencies can increase the set of beneficial reformulations of a database. Consider an example:
\vspace{-0.2cm}
\begin{example}
\label{intro-first-example}
Let a query $Q$ be defined on a database with schema $\{ S(A,B), \ T(C,D) \}$ as
\vspace{-0.4cm}
\begin{tabbing}
This is my \= kill mark \kill
$q(X,Y) \ :- \ s(X,Y), \ s(X,a), \ t(Y,a)$.
\end{tabbing}
\vspace{-0.4cm}
Consider a view $V$,
\vspace{-0.4cm}
\begin{tabbing}
This is m \= kill mark \kill
$v(X,W) \ :- \ s(X,a), \ t(a,W)$.
\end{tabbing}
\vspace{-0.4cm}
Query $Q$ --- but not view $V$ --- {\it has self-joins}, that is, the definition of $Q$ but not of $V$ has multiple literals with the same relation name. It can be shown~\cite{LevyMSS95} that in the absence of dependencies, $V$ cannot be used to equivalently rewrite $Q$. At the same time, suppose the database satisfies a functional dependency $\sigma$,
\vspace{-0.4cm}
\begin{tabbing}
This is \= my kill mark \kill
$\sigma: \ \forall X,Y,Z \ (s(X,Y) \ \wedge \ s(X,Z) \rightarrow (Y = Z)).$
\end{tabbing}  
\vspace{-0.4cm}
This dependency means that whenever two tuples in relation $S$ agree on the value of the first attribute $A$, they also agree on the value of the second attribute $B$ of $S$.

On all databases satisfying the dependency $\sigma$, the query $Q$ can be equivalently rewritten\footnote{We assume set semantics~\cite{ChaudhuriV93} for query evaluation.} using the view $V$, as follows:
\vspace{-0.4cm}
\begin{tabbing}
This is my \= kill mark \kill
$q(X,a) \ :- \ v(X,a)$.
\end{tabbing}
\vspace{-0.4cm}
The reformulation is optimal on all databases satisfying $\sigma$, as the materialized view $V$ precomputes an exact answer to $Q$. 
\end{example}
\vspace{-0.2cm}

In this paper we enhance the results of~\cite{ChirkovaG00} to deal with the additional complexities that arise in presence of dependencies. The problem we consider is as follows: given a set of queries, a set of dependencies, and a storage limit, is it possible to efficiently generate reformulations that satisfy the storage limit and minimize the total costs of evaluating the queries, in the presence of the dependencies. We look at this problem for conjunctive queries, views, and rewritings on finite databases in presence of several types of dependencies, including functional and inclusion dependencies. Our results are applicable in data warehousing and query optimization. Our contributions are as follows:
\vspace{-0.2cm}
\begin{itemize}
\vspace{-0.2cm}
	\item we give a new algorithm and tighter complexity results for database reformulation in the absence of dependencies, for queries without self-joins (Section~\ref{cgalg-subsection});
\vspace{-0.2cm}
	\item we give complexity results and algorithms for database reformulation in presence of dependencies, based on the chase technique~\cite{AbiteboulHV95} for incorporating dependencies into query definitions (Section~\ref{dep-chase-section});
\vspace{-0.2cm}
	\item we introduce an {\it unchase} technique for reducing the problem of query equivalence under dependencies to query equivalence in the absence of dependencies, {\it without} increasing query size (Section~\ref{unchase-section}); 
\vspace{-0.2cm}
	\item we show that we can reduce the complexity of database reformulation {\it and} cover larger classes of dependencies by basing the reformulation algorithm on the unchase approach (Section~\ref{unchase-section}).
\end{itemize}
\vspace{-0.4cm}

After covering related work in the remainder of this section, we give basic definitions and formal problem statement in Section~\ref{prelim-section}. We then present complexity results and algorithms for database reformulation: Section~\ref{dep-chase-section} describes an approach based on chase, and Section~\ref{unchase-section} discusses our unchase technique. We conclude and discuss future work in Section~\ref{conclus-section}.

\vspace{-0.4cm}
\subsection*{Related work}
\vspace{-0.3cm}

\label{relwork-section}

Studies of dependencies have been motivated by the goal of good database schema design; interestingly, they have also contributed to basic research in mathematical logic. The study of dependency theory began with the introduction of functional dependencies in~\cite{Codd72}; inclusion dependencies were first identified in~\cite{CasanovaFP82}. The topic of queries defined over databases that satisfy dependencies was initiated in~\cite{AhoSU79a,AhoSU79b}. Containment in the presence of inclusion dependencies has been examined in~\cite{KanellakisCV83,JohnsonK84}.  For surveys and references on data dependencies, see~\cite{FaginV84,Kanellakis90,AbiteboulHV95}. 

An important technique named {\it chase} grew out of the algorithm of~\cite{AhoBU79} for testing lossless joins. The chase can be further extended into a semidecision procedure for embedded-dependency implication and an exponential decision procedure for full dependency implication, see~\cite{BeeriV84,BeeriV84b}. In its most general form, chase is similar to resolution with paramodulation. See~\cite{DeutschDiss,DeutschLN05} and references therein for applications of chase to answering queries equivalently using views. 

Conjunctive queries~\cite{ChandraM77,AhoSU79a,AhoSU79b} form a large and well-studied class of queries that contains a large proportion of those questions one might wish to ask in practice. When there are no dependencies to consider, or when there are only functional dependencies, it has been shown that the containment, equivalence, and minimization problems are all NP-complete~\cite{ChandraM77}. These results should not be viewed as negative, especially for problems concerned with query optimization, since queries are typically much smaller than the databases on which they are asked, and queries may be applied repeatedly over time~\cite{JohnsonK84}. 

References to view selection can be found in~\cite{Halevy01,ChirkovaHS02,AfratiC05}. To the best of our knowledge, the results presented here are the first results on view selection in presence of dependencies.

\vspace{-0.5cm}
\section{Preliminaries}
\vspace{-0.4cm}

\label{prelim-section}

In this section we provide definitions and technical background for our framework, using in part the materials in~\cite{Kanellakis90,AbiteboulHV95}.

\vspace{-0.5cm}
\subsection{Basic definitions}
\vspace{-0.3cm}

A {\em relational database} is a finite collection of stored relations. Each relation $R$ is a finite set of tuples, where each tuple is a list of values of the attributes in the {\em relation schema} of $R$. We consider select-project-join SQL queries with equality comparisons, a.k.a. {\em safe conjunctive queries}. A conjunctive query is a rule of the form: $Q: \ \ q(\bar{X})$ $\leftarrow \ e_1(\bar{X}_1), \dots, e_n(\bar{X}_n)$, where $e_1, \dots\, e_n$ are names of database relations and $\bar{X}, \bar{X}_1, \dots, \bar{X}_n$ are vectors of variables. A query $Q$ {\it has self-joins} if at least two different atoms $e_i(\bar{X}_i)$, $e_j(\bar{X}_j)$ in the body of $Q$ have the same relation name. The variables in $\bar{X}$ are called {\em head} or {\em distinguished variables} of $Q$,
whereas the variables in $\bar{X}_i$ are called {\em body variables} of $Q$. A query is {\it safe} if $\bar{X} \subseteq \bigcup_{i = 1}^n \bar{X}_i$. 

\vspace{-0.5cm}
\subsection{Dependencies and chase}
\vspace{-0.3cm}
\label{chase-prelim-section}

A {\it dependency} over a database schema $\cal S$ is a sentence in some logical formalism over $\cal S$. We consider {\it tuple-generating dependencies (tgds)} and {\it equality-generating dependencies (egds)}~\cite{BeeriV84}. A tgd is of the form 
$\forall \ \bar{x} \ (\phi(\bar{x}) \rightarrow \exists \ \bar{y} \ \psi(\bar{x},\bar{y})),$ 
and an egd is of the form 
$\forall \ \bar{x} \ (\phi(\bar{x}) \rightarrow (x_i = x_j)).$ 
Here, $\bar{x} = x_1,\ldots,x_k$, $\bar{y} = y_1,\ldots,y_m$, and each of $x_i, x_j$ is an element in $\bar{x}$. In addition, we consider {\it consistency constraints} of the form 
$\forall \ \bar{x} \ (\phi(\bar{x}) \rightarrow \ false).$ 
In this paper, we consider {\it conjunctive} egd's, tgd's, and consistency constraints, that is, in all the dependencies we consider, $\phi(\bar{x})$ is a conjunction of relational atoms, and $\psi(\bar{x},\bar{y})$ (in the tgd's) is a single relational atom. We refer to conjunctive egd's as {\it functional dependencies} (fds), and distinguish between two types of conjunctive tgd's: In {\it value-preserving tgd's}, $\bar{y}$ in the right-hand side in empty, and in {\it value-generating tgd's}, $\bar{y}$ contains at least one variable name. In many results in this paper, we focus on a special case of conjunctive tgd's called {\it inclusion dependencies} (ids), which have just one relational atom in the left-hand side. Inclusion dependencies may be value preserving or value generating. We will use a shorthand notation, in which quantifiers are not used where clear from context. For ids we will also use the notation $r[\bar{x}] \subseteq s[\bar{x}]$, which is equivalent to $\forall \ \bar{x}, \ \bar{z} \ (r(\bar{x},\bar{z}) \rightarrow \exists \ \bar{y} \ s(\bar{x},\bar{y})).$

A set $\Sigma$ of ids is {\it acyclic} if there is no sequence $r_i[\bar{x}_i] \subseteq s_i[\bar{x}_i]$ ($i \ \epsilon \ [1,\ldots n]$) of ids in $\Sigma$ where for $i \ \epsilon \ [1,\ldots n]$, $r_{i+1} = s_i$ for $i \ \epsilon \ [1,\ldots, n-1]$, and $r_1 = s_n$. A family $\Sigma$ of dependencies {\it has acyclic ids} if the set of ids in $\Sigma$ is acyclic~\cite{AbiteboulHV95}. We define {\it acyclic tgds} as follows: A set $\Sigma$ of tgds is {\it acyclic} if there is no sequence $\sigma_{1}, \sigma_2, \ldots, \sigma_n$ of tgds of $\Sigma$, $\sigma_i: \ r_{i1}(\bar{x}_{i1}) \ \wedge \ \ldots \ \wedge \ r_{ik}(\bar{x}_{ik})  \rightarrow s_i(\bar{y}_i)$ ($i \ \epsilon \ [1,\ldots n]$) of tgds in $\Sigma$ where for $i \ \epsilon \ [1,\ldots n]$, the left-hand side of $\sigma_{i+1}$ includes the relation name for the right-hand side of $\sigma_i$, for $i \ \epsilon \ [1,\ldots n-1]$, and the left-hand side of $\sigma_1$ includes the relation name for the right-hand side of $\sigma_n$. A set $\Sigma$ of $n$ acyclic tgds is {\it strongly acyclic} if there exists a sequence $\sigma_{1}, \sigma_2, \ldots, \sigma_n$ of {\it} tgds of $\Sigma$, such that for $i \ \epsilon \ [1,\ldots, n-1]$ and for all $k > 0$ such that $i + k \leq n$, the right-hand side of $\sigma_{i+k}$ does not include any relation name in the left-hand side of $\sigma_i$. A family $\Sigma$ of dependencies {\it has (strongly) acyclic tgds} if the set of tgds in $\Sigma$ is (strongly) acyclic.

We denote the left-hand side of a dependency (or the body of a query) by $A$. An {\it assignment} $\gamma$ for $A$ is a mapping of the variables appearing in $A$ to constants, and of the constants in $A$ to themselves. Assignments are naturally extended to tuples and atoms; for instance, 
for a tuple of variables $\bar{s} = (s_1,\ldots,s_k)$ we let $\gamma\bar{s}$ denote the tuple
$(\gamma(s_1),\ldots,\gamma(s_k))$. {\it Satisfaction} of atoms by an assignment w.r.t a database is defined as follows: $p_i(\gamma \bar{s})$ is satisfied if the tuple $\gamma \bar{s}$ is in the relation that corresponds to the predicate of $p_i$. This definition is naturally extended to that of satisfaction of conjunctions of atoms. An {\it answer to a safe query} $Q$ with head $q(\bar{x})$ and body $A$ on a database $\cal D$ is the set of all tuples $\gamma(\bar{x})$ such that $\gamma$ is a satisfying assignment for $A$ on $\cal D$.

A database $\cal D$ {\it satisfies a set of dependencies} $\Sigma$ if, for each dependency $\sigma$ in $\Sigma$ and for all satisfying assignments $\gamma$ of the left-hand side of $\sigma$ w.r.t. $\cal D$, $\sigma$ evaluates to true. (For value-generating tgd's $\sigma$, we additionally require that we can extend each $\gamma$ in such a way that the right-hand side of $\sigma$ evaluates to true.) For a given set $\Sigma$ of dependencies and conjunctive queries $Q_1$ and $Q_2$, $Q_1$ is {\em contained} in $Q_2$ under $\Sigma$, denoted by $Q_1 \sqsubseteq_{\Sigma} Q_2$, if for any database $\cal D$ that satisfies $\Sigma$, the answer to $Q_1$ on $\cal D$ is a subset of the answer to $Q_2$ on $\cal D$. Two queries are {\em equivalent under $\Sigma$} if they are contained in each other under $\Sigma$. Query containment and equivalence in the absence of dependencies is defined as above for the case $\Sigma = \phi$ (empty set). 

In this paper we use the following results of~\cite{ChandraM77} for conjunctive queries. In the absence of dependencies, a query $Q_1$ is contained in $Q_2$ if and only if there exists a {\it containment mapping} from $Q_2$ to $Q_1$, that is, a homomorphism from the variables of $Q_2$ to the variables and constants of $Q_1$, such that (1) each atom in the body of $Q_2$ is mapped into some atom in the body of $Q_1$, and (2) the head of $Q_2$ is mapped into the head of $Q_1$. For a query $Q$, its {\it minimized} version is an equivalent query $Q'$ with a minimum number of subgoals, which can be obtained via repeated applications of containment mappings. Two queries are equivalent if and only if their minimized versions are isomorphic.

It is easy to show the following:
\vspace{-0.2cm}
\begin{proposition}
Given a database schema $\cal S$, queries $Q_1$ and $Q_2$ defined on $\cal S$, and a set $\Sigma$ of dependencies on $\cal S$, if $Q_1$ is contained in $Q_2$ in the absence of dependencies, $Q_1 \sqsubseteq Q_2$, then $Q_1$ is contained in $Q_2$ under $\Sigma$, $Q_1 \sqsubseteq_{\Sigma} Q_2$.
\end{proposition}
\vspace{-0.2cm}

The {\it chase} is a process that, given dependencies $\Sigma$, transforms a query $Q$ into a query $Q'$ such that $Q \equiv_{\Sigma} Q'$. A {\it chase sequence} of a conjunctive query $Q: q \ :- \ A$ by a set of dependencies $\Sigma$ is a (possibly infinite) sequence of conjunctive queries $(q_0,A_0), (q_1,A_1), \ldots, (q_i,A_i), \ldots, $ 
where $q_0 = q$ and $A_0 = A$, and for each $i \geq 0$, the query $(q_{i+1},A_{i+1})$ is the result of applying some dependency in $\Sigma$ to the query $(q_i,A_i)$. We can apply a dependency to a query if there is a satisfying assignment $\gamma$ of the left-hand side of the dependency w.r.t. the body of the query. For fds, the chase rule is to consistently rename query variables according to the equality in the right-hand side of the fd. For inds, the chase rule~\cite{JohnsonK84} adds to a partial chase result $(q_i,A_i)$ a subgoal that matches the right-hand side $p(\bar{x})$ of the ind, provided no existing subgoal in $(q_i,A_i)$ matches $p(\bar{x})$. This rule is extended to tgds in a natural way. The chase sequence is {\it terminal} if (1) it is finite, and (2) no dependency in $\Sigma$ can be applied to the last element in the sequence. The {\it result} of a terminal chase sequence is its last element $(q_n,A_n)$, written in query form as $Q': q_n \ :- \ A_n$.

\vspace{-0.4cm}
\begin{definition}{Chase}
For a query $Q$ and a set of dependencies $\Sigma$, the {\it chase} of $Q$ by $\Sigma$, denoted $chase_{\Sigma}(Q)$, is the result of any terminal chasing sequence of $Q$ by $\Sigma$.
\end{definition}
\vspace{-0.4cm}

Given a query $Q$ and dependencies $\Sigma$, we compute $chase_{\Sigma}(Q)$ by picking the dependencies in $\Sigma$ in some arbitrary order and applying them to $Q$. Importantly, the chase is determined by the {\it semantics}, rather than the {\it syntax}, of the dependencies in $\Sigma$. Let $\Sigma$ and $\Sigma'$ be two sets of dependencies over schema $\cal S$. If $\Sigma \equiv \Sigma'$\footnote{$\Sigma \equiv \Sigma'$ if $\Sigma \models \Sigma'$ and $\Sigma' \models \Sigma$.}, then $chase_{\Sigma}(Q)$ and $chase_{\Sigma'}(Q)$ coincide for any query $Q$. 

The following result has been shown for sets of functional dependencies in~\cite{AbiteboulHV95}; we have extended it to sets of any dependencies considered in this paper.

\vspace{-0.4cm}
\begin{theorem}
\label{chase-theorem}
Given conjunctive queries $Q_1$, $Q_2$ and a set $\Sigma$ of fds, conjunctive consistency constraints, and conjunctive tgds.
\vspace{-0.3cm}
\begin{enumerate}
	\item $Q_1 \sqsubseteq_{\Sigma} Q_2$ iff $chase_{\Sigma}(Q_1) \sqsubseteq chase_{\Sigma}(Q_2)$ in the absence of any constraints.
\vspace{-0.3cm}
	\item $Q_1 \equiv_{\Sigma} Q_2$ iff $chase_{\Sigma}(Q_1) \equiv chase_{\Sigma}(Q_2)$ in the absence of any constraints.
\end{enumerate}
\vspace{-0.9cm}
\end{theorem}
\vspace{-0.4cm}

\vspace{-0.4cm}
\subsection{Views and database reformulation}
\vspace{-0.3cm}

A {\em view} refers to a named query. A view is said to be {\em materialized} if its answer is stored in the database. Let $\cal V$ be a set of views defined on a database schema $\cal
S$, and  $\cal D$ be a database with schema $\cal S$; by
${\cal D}_{\cal V}$ we denote the database obtained by computing
all the view relations in $\cal V$ on $\cal D$. Let $Q$ be a query
defined on $\cal S$, and $\cal V$ be a set of views defined on
$\cal S$. A query $R$ is a {\em rewriting} {\em of $Q$
using $\cal V$} if all atoms in the body of $R$ are vie
predicates defined in $\cal V$.

The {\em expansion} $R^{exp}$ of a rewriting $R$ of a query $Q$ on a set of
views $\cal V$ is obtained from $R$ by
replacing all the view atoms in the body of $R$ by their
definitions in terms of the base relations. A rewriting $R$ of a query $Q$ on a set of views $\cal V$ is an {\em equivalent rewriting} of $Q$ under $\Sigma$ if for every database $\cal D$ that satisfies $\Sigma$, $Q(\cal D) = R({\cal D}_{\cal V})$.

We consider the following {\it database-reformulation} problem: Given a set of conjunctive queries $\cal Q$ on stored relations, a fixed database instance $\cal D$ that satisfies a set of dependencies $\Sigma$, and a storage limit $L$, we want to find and precompute offline a set of views on the stored relations. A set of views $\cal V$ is {\it admissible} for $({\cal Q}, {\cal D}, \Sigma, L)$ if (1) $\cal V$ provides an equivalent rewriting for each query in $\cal Q$ under $\Sigma$, and (2) the total size of the relations for $\cal V$ on $\cal D$ does not exceed the storage limit $L$. (The {\it size} of a relation is the number of bytes used to store the relation.) Among such admissible sets of views, our goal is to find a {\it beneficial} (or {\it optimal}) viewset, that is, a set of views whose use in rewritings of the queries in $\cal Q$ reduces (minimizes) the sum of evaluation costs of these queries on the database $\cal D$ satisfying the dependencies $\Sigma$. For query-evaluation costs, we consider {\it size-monotonic} cost models, where (1) query costs are computed using the sizes of the contributing relations, and (2) whenever a relation in a query expression is replaced by another relation of at most the same size, the cost of evaluating the new expression is at most the cost of evaluating the original expression. All the common cost models in the literature are size-monotonic.

\vspace{-0.4cm}
\begin{definition}{Database reformulation}
For a problem input ${\cal I} = ({\cal Q}, {\cal D}, \Sigma, L)$, a {\em beneficial} ({\em optimal}) {\em
viewset} is a set of views $\cal V$ defined on $\cal S$, such that:  
(1) $\cal V$ is an admissible viewset for $\cal I$, and
(2) $\cal V$ reduces (minimizes) the total cost of evaluating the queries in $\cal Q$ on the database ${\cal D}_{\cal V}$.
\end{definition}
\vspace{-0.4cm}

We consider this problem in relational databases for conjunctive queries, views, and rewritings. We assume that filtering views are not used in query rewritings.\footnote{In an equivalent rewriting $R$ of a query $Q$, a view $V$ is a {\em filtering view} if the result of removing the literal for $V$ from $R$ is still an equivalent rewriting of $Q$.} In some results we additionally assume that input queries do not have self-joins.
We use these simplifying assumptions to do an initial study of the structure of the database-reformulation problem under dependencies. It is known that when these assumptions do not hold, the problem has a triply exponential upper bound and a singly exponential lower bound even in the absence of dependencies~\cite{ChirkovaHS02}. The database-reformulation problem is in NP in the absence of dependencies when input queries do not have self-joins and when filtering views are not used~\cite{AfratiCGP05}.

\vspace{-0.5cm}
\subsection{The {\tt cgalg} algorithm~\cite{ChirkovaG00}}
\vspace{-0.3cm}

\label{cgalg-subsection}

We now outline an algorithm for generating beneficial reformulations for the case where the set of dependencies $\Sigma$ is empty and $\cal Q$ comprises a single query $Q$~\cite{ChirkovaG00}. For each beneficial reformulation (viewset) $\cal V$ for a problem input $\cal I$, this algorithm generates at least one beneficial reformulation (viewset) ${\cal V}'$ that reduces the costs of the input query workload at least as much as $\cal V$ and satisfies the same storage limit. We say that the algorithm produces the {\it best} beneficial database reformulations.

\vspace{-0.1cm}
{\small
\noindent
{\bf Procedure} {\sl cgalg}. \\
Input: query $Q$, database $\cal D$, storage limit $L$.\\
Output: $R_{opt}$, optimal equiv. rewriting of $Q$ on $\cal D$.\\
1 Begin: \\
2 \hspace*{.4cm}minimize $Q$ to obtain a query $Q'$; \\
3 \hspace*{.4cm}set $R_{opt}$ to $Q'$;\\
4 \hspace*{.4cm}set the cost $C_{opt}$ of $R_{opt}$ to $C(Q')$;\\
5 \hspace*{.4cm}find all views $\cal V$ whose body is a subset \\ 
\hspace*{.6cm} of subgoals of $Q'$; \\
6 \hspace*{.4cm}for each subset $\cal W$ of $\cal V$ such that \\
\hspace*{.6cm} $\Sigma_{W \ \epsilon \ {\cal W}} size(W,{\cal D}) \leq L$ do: \\
7 \hspace*{.4cm}begin:\\
8   \hspace*{1.1cm}find a rewriting $R$ of $Q'$ using $\cal W$;\\
9   \hspace*{1.1cm}construct the expansion $R^{exp}$ of $R$;\\
10   \hspace*{1cm}if there exists a containment mapping \\
\hspace*{1.3cm} from $Q'$ to $R^{exp}$ then:\\
11     \hspace*{1.6cm}if the cost $C(R,{\cal D},{\cal O})$ of answering $Q'$ \\
\hspace*{1.9cm} on $\cal D$ using $R$ is less than $C_{opt}$ \\
12     \hspace*{1.6cm}then begin: \\
13 \hspace*{2.3cm}  $R_{opt} \ := \ R$; \\
14 \hspace*{2.3cm}  $C_{opt} \ := \ C(R,{\cal D},{\cal O})$; \\
15 \hspace*{1.6cm}  end; \\
16 \hspace*{.4cm}end;\\
17 \hspace*{.4cm}return $R_{opt}$.\\
18 End.\\
}
\vspace{-0.4cm}

A view-size oracle $\cal O$ instantaneously gives the size of any relation defined on the database $\cal D$; we assume that for a rewriting $R$ in terms of views and for a fixed size-monotonic cost model for query evaluation, the time required to obtain the cost $C(R,{\cal D},{\cal O})$ of evaluating $R$ in terms of the relations for the views on $\cal D$ is negligible when using the oracle $\cal O$. In practice, the view sizes and costs of answering $Q$ on $\cal D$ using $R$ can be estimated via standard formulas used in query optimizers in database-management systems. It is easy to see how the {\tt cgalg} algorithm can be extended to problem inputs with non-singleton query workloads.\footnote{When queries have no self-joins and each view in $\cal V$ is used exactly once in the rewriting of exactly one query in $\cal Q$~\cite{AfratiCGP05}, {\tt cgalg} can look for views for each workload query separately even when $\Sigma$ is not empty.}

\vspace{-0.4cm}
\begin{proposition}{~\cite{ChirkovaG00,AfratiCGP05}}
\label{cgalg-soundness}
Given $\Sigma = \phi$ and provided that all view atoms in all rewritings have different relation names and that filtering views are not used in query rewritings, the algorithm {\tt cgalg} is sound for problem inputs with workloads of arbitrary conjunctive queries and is complete for problem inputs with workloads of conjunctive queries without self-joins. The decision version of the problem of finding optimal reformulations is NP complete.
\end{proposition}
\vspace{-0.4cm}
In general, the algorithm is not complete (i.e., is not guaranteed to produce an optimal reformulation) because some optimal rewritings may use self-joins of view literals~\cite{ChirkovaHS02}.

\vspace{-0.4cm}
\begin{proposition}
Under the assumptions of Proposition~\ref{cgalg-soundness} and assuming that a view-size oracle $\cal O$ and a size-monotonic cost model for query evaluation are given, the runtime of {\tt cgalg} is $\Theta(2^m)$, where $m$ is the total number of subgoals of the queries in the input workload $\cal Q$.
\end{proposition}
\vspace{-0.4cm}
Intuitively, under the assumptions of Proposition~\ref{cgalg-soundness}, {\tt cgalg} will generate all beneficial reformulations if it generates only viewsets that have up to $m$ views~\cite{AfratiCGP05}. Note that the step of generating a rewriting given a subset ${\cal W}$ of the set $\cal V$ of views takes constant time in the size of the subset ${\cal W}$~\cite{AfratiLU01}.

\vspace{-0.5cm}
\section{Dependencies and Chase}
\vspace{-0.3cm}

\label{dep-chase-section}

In this section and in Section~\ref{unchase-section}, we consider the database-reformulation problem for workloads of conjunctive queries under a nonempty set of dependencies $\Sigma$. In this section our focus is on using chase to extend the {\tt cgalg} algorithm (Section~\ref{cgalg-subsection}) to database reformulation in presence of dependencies.

We first observe that the straightforward approach to finding all useful views and rewritings does not really work. Given a query $Q$ and a set of dependencies $\Sigma$, we can use Theorem~\ref{chase-theorem} to reduce the problem of finding rewriting expansions that are equivalent to $Q$ under $\Sigma$ to the problem of finding rewriting expansions whose terminal chase result (under $\Sigma$) is equivalent, {\it in the absence of} $\Sigma$, to the terminal chase result $Q_c$ of $Q$ under $\Sigma$. Even if $Q_c$ is unique and finite, the number of queries that are equivalent to $Q_c$ is infinite~\cite{ChandraM77}, and the number of {\it all} beneficial views and rewritings can be infinite~\cite{ChirkovaG00}. In this section we use chase to extend the approach of~\cite{ChirkovaG00} of generating the best (rather than all) beneficial viewsets using the {\tt cgalg} algorithm.

\vspace{-0.5cm}
\subsection{Consistency constraints}
\vspace{-0.3cm}

We first obtain that consistency constraints do not generate new views.

\vspace{-0.4cm}
\begin{theorem}
Let $\cal I$ be a problem input where all dependencies in $\Sigma$ are consistency constraints. Then an optimal set of views $\cal V$ for $\cal I$ can be found by finding an optimal set of views for the problem input that is obtained by removing all dependencies from $\cal I$.
\end{theorem}
\vspace{-0.4cm}

A corollary of this result is that if at least one consistency constraint is combined with any number of fds and tgds, then the database-reformulation output is the same as for a problem input where all the consistency constraints are removed.

\vspace{-0.5cm}
\subsection{Functional dependencies}
\vspace{-0.3cm}

\label{fd-chase-subsection}

As we saw in Example~\ref{intro-first-example} in Section~\ref{intro-section}, unlike consistency constraints, fds can generate new beneficial reformulations. 

\vspace{-0.4cm}
\begin{lemma}{~\cite{AbiteboulHV95}} 
Let $\Sigma$ be a set of fds; for any query $Q$, let $Q' = chase_{\Sigma}(Q)$. Then (1) $Q'$ is unique up to variable renamings, and (2) the size of the minimized version of $Q'$ does not exceed the size of the minimized version of $Q$.
\end{lemma}
\vspace{-0.7cm}

\begin{theorem}
Algorithm {\tt cgalg$(\{ chase_{\Sigma}(Q) \},$} {\tt ${\cal D},L)$} produces an optimal reformulation of a problem input $\cal I$ where ${\cal Q} = \{ Q \}$ and where all dependencies in $\Sigma$ are fds, provided that all queries in the workload $\{ chase_{\Sigma}(Q) \}$ have no self-joins. 
\end{theorem}
\vspace{-0.4cm}
Note that to produce an optimal reformulation, we only need to consider the {\it terminal} chase result of each query in the workload $\cal Q$. The complexity of {\tt cgalg} here does not exceed the complexity of {\tt cgalg} for the same problem input in the absence of dependencies; note that the original queries may have self-joins (see Example~\ref{intro-first-example}).

\vspace{-0.5cm}
\subsection{Conjunctive tgds}
\vspace{-0.3cm}

We now consider problem inputs whose dependency sets $\Sigma$ contain acyclic sets of conjunctive tgds. We first consider the case where all tgds are ids. 

\vspace{-0.4cm}
\begin{proposition}{~\cite{AbiteboulHV95}}
Let $Q$ be a query and $\Sigma$ a set of fds and acyclic ids. Then each chasing sequence of $Q$ by $\Sigma$ terminates after an exponentially bounded number of steps.
\end{proposition}
\vspace{-0.7cm}

\begin{proposition}
\label{jk-proposition}
Let $\Sigma$ be a of fds $\Sigma[F]$ and acyclic ids $\Sigma[I]$, $\Sigma[F] \cup \Sigma[I] = \Sigma$. Then for all conjunctive queries $Q$, $chase_{\Sigma}(Q) = chase_{{\Sigma}[I]}(chase_{{\Sigma}[F]}(Q))$.
\end{proposition}
\vspace{-0.4cm}
This result extends the result of~\cite{JohnsonK84} for a special class of sets of fds and ``key-based" ids; to obtain the extension, we use the observation that the chase rule for ids in~\cite{JohnsonK84} (which we also use) does not add to the partial chase result $Q_{c,p}$ the right-hand side of a qualifying id if a matching subgoal is already in $Q_{c,p}$. In extending the result to acyclic tgds, the subtlety is that (part of) the left-hand side of a tgd can match the left-hand side of an fd in the same set of dependencies, which would cause Proposition~\ref{jk-proposition} to be violated. (For instance, $\Sigma$ can include a tgd $s(X,Y) \ \wedge \ s(X,Z) \rightarrow p(X,Z)$ and an fd $s(X,Y) \ \wedge \ s(X,Z) \rightarrow Y = Z$.) We obtain the result of Proposition~\ref{jk-proposition} for sets of dependencies $\Sigma$ that have been preprocessed, by applying each fd in $\Sigma$ to the left-hand side of each tgd in $\Sigma$.

\vspace{-0.4cm}
\begin{theorem}
\label{main-chase-id-theorem}
Algorithm {\tt cgalg$(\{ chase_{\Sigma}(Q) \},$} {\tt ${\cal D},L)$} is sound for problem inputs $\cal I$ where ${\cal Q} = \{ Q \}$ and where $\Sigma$ is a set of fds and acyclic tgds. The algorithm is complete for such inputs if queries $chase_\Sigma (Q)$ have no self-joins.
\end{theorem}
\vspace{-0.4cm}

\vspace{-0.5cm}
\section{Reducing the Complexity by Unchase}
\vspace{-0.3cm}

\label{unchase-section}

In Section~\ref{dep-chase-section} we saw that we can obtain the best beneficial reformulations for a workload of conjunctive queries in presence of consistency constraints, fds, and acyclic tgds, either separately or in combination, by using the {\tt cgalg} algorithm on the terminal chase results of the workload queries. At the same time, the restrictions on this approach are rather strong. First, the terminal chase result of each query cannot have self-joins if we want to obtain {\it optimal} reformulations. Second, as shown in Section~\ref{cgalg-subsection}, the complexity of {\tt cgalg} is exponential in the size of the queries to which {\tt cgalg} is applied, that is, to the terminal chase results of the workload queries. 

We now give an example where the terminal chase result of a query under acyclic ids (1) has self-joins, and (2) is of size exponential in the size of the query. Thus, the {\tt cgalg} approach of Section~\ref{dep-chase-section} is not guaranteed to produce optimal reformulations in this case,  and the cost of using the approach to produce some beneficial reformulations would be prohibitive even for simple queries. However, in this section we give a modified {\tt cgalg} approach that is applicable to the problem input of this example and to other cases, including problem inputs where the terminal chase results of the input queries under the input dependencies are infinite in size.

\vspace{-0.4cm}
\begin{example}
\label{expon-example}
On a database schema ${\cal S} = \{ P_1(A_1,B_1), \ P_2(A_2,B_2), \ \ldots, \ P_m(A_m,B_m) \}$, consider a query $Q$ with a single subgoal $p_1$:
\vspace{-0.4cm}
\begin{tabbing}
This is my \= kill mark \kill
$q(X,Y) \ :- \ p_1(X,Y)$.
\end{tabbing}
\vspace{-0.4cm}
Suppose the database schema $\cal S$ satisfies a set $\Sigma$ of acyclic ids of the following form:
\vspace{-0.4cm}
\begin{tabbing}
$\sigma^{(1)}_{i,j}: \ p_i(X,Y) \rightarrow p_{j}(Z,X)$ \\
$\sigma^{(2)}_{i,j}: \ p_i(X,Y) \rightarrow p_{j}(Y,W)$ 
\end{tabbing}
\vspace{-0.4cm}
$\Sigma$ has one id $\sigma^{(1)}_{i,j}$ and one id $\sigma^{(2)}_{i,j}$ for each pair $(i,j)$, where $i \ \epsilon \ \{ 1,\ldots,m-1 \}$ and $j \ \epsilon \ \{ i+1,\ldots,m \}$ ($i < j$ in each pair). Thus, the number of dependencies in $\Sigma$ is quadratic in $m$. 

We show one partial chase result of the query $Q$ under dependencies $\Sigma$, for $m \geq 2$:
\vspace{-0.4cm}
\begin{tabbing}
This is my \= kill mark \kill
$q'(X,Y) \ :- \ p_1(X,Y), \ p_2(Z_1,X), \ p_2(Y,Z_2)$.
\end{tabbing}
\vspace{-0.4cm}
This query $Q'$ is the result of applying to $Q$ dependencies $\sigma^{(1)}_{1,2}$ and $\sigma^{(2)}_{1,2}$. 

For the terminal result $Q_c$ of chasing the query $Q$ under the ids $\Sigma$, we can show that the size of $Q_c$ is exponential in the size of $Q$ and $\Sigma$.
\end{example}
\vspace{-0.4cm}

For the problem input in this example, the cost of using {\tt cgalg} of Section~\ref{dep-chase-section} is doubly exponential in the size of the query $Q$, and the problem of finding beneficial reformulations has an exponential-size lower bound, just because we need to output views that cover all the subgoals of this exponential-size terminal chase result. 
Thus, the terminal chase result of a query under acyclic ids can have an exponential number of views even for (1) nonfiltering views only, and (2) no self-joins in input queries (cf.~\cite{AfratiCGP05}). 

\vspace{-0.5cm}
\subsection{Unchase for ids and tgds}
\vspace{-0.3cm}

The idea we outline in this section is to apply our reformulation algorithm to those versions of the input queries that have all ``derived" subgoals removed. Thus, our approach is to (1) apply ``unchase" to all the input queries under the input dependencies, and then to (2) apply {\tt cgalg} to the results of the unchase. 

We first define unchase for sets of ids only: Given a finite-size query $Q$ and an id $\sigma$, an {\it unchase step} on $(Q,\sigma)$ is to remove from $Q$ a subgoal $s$ that is the image, under some homomorphism $\mu$, of the right-hand side $r$ of $\sigma$, provided that two conditions are satisfied. First, the homomorphism $\mu$ can be extended to map the left-hand side of $\sigma$ into some subgoal of $Q$ other than $s$. Second, for each {\it free} argument $Y$ of $r$, $\mu(Y)$ in $s$ (1) is a variable rather than a constant, (2) is a nondistinguished variable of $Q$, and (3) does not occur in any subgoal of $Q$ except $s$. For instance, if we apply the id $\sigma^{(1)}_{1,2}:  \ p_1(X,Y) \rightarrow p_{2}(Z,X)$ to query $Q'$ in Example~\ref{expon-example}, we will obtain a query 
$q''(X,Y) \ :- \ p_1(X,Y), \ p_2(Y,Z_2).$

For a query $Q$ and for a set of ids $\Sigma$, we denote by $Q_{u,\Sigma}$ the terminal unchase result of $Q$ under $\Sigma$. Note that unchase under ids terminates in finite time, as each successful unchase step removes a subgoal from the current partial unchase result. We obtain the following uniqueness result for unchase under ids:
\vspace{-0.4cm}
\begin{lemma}
\label{lemma-x}
For a conjunctive query $Q$, for a set of dependencies $\Sigma$ that has ids only, and for any finite-size (either partial or terminal) chase result $Q'$ of $Q$ under $\Sigma$, $Q_{u,\Sigma}$ is equivalent to $Q'_{u,\Sigma}$ in the absence of dependencies.
\end{lemma}
\vspace{-0.4cm}
It follows~\cite{ChandraM77} that the result of minimizing $Q_{u,\Sigma}$ is isomorphic to the result of minimizing $Q'_{u,\Sigma}$. Note that in Lemma~\ref{lemma-x} we do not require id acyclicity, and thus the result applies to problem inputs with sets of cyclic ids, such as $\{ p(X,Y) \rightarrow p(Y,Z) \}$. We have also extended the result of Lemma~\ref{lemma-x} to sets of strongly acyclic tgds; the unchase rule for tgds is analogous to that for ids. (We require strong acyclicity in the proof to ensure that all tgds can be applied in the unchase process.)

\vspace{-0.5cm}
\subsection{Unchase in presence of fds}
\vspace{-0.3cm}

Using Lemma~\ref{lemma-x}, we can show that {\tt cgalg} can be applied to the problem input of Example~\ref{expon-example} to obtain an optimal reformulation from just the terminal unchase result (which is $Q$ itself) of the query $Q$ under the set $\Sigma$ of ids. However, we can extend the unchase/{\tt cgalg} approach to combinations of ids (or of strongly acyclic tgds) with fds. We first note that if we try to unchase a query using fds only, the unchase process will not terminate in finite time:
\vspace{-0.2cm}
\begin{example}
For a query 
\vspace{-0.4cm}
\begin{tabbing}
This is my \= kill mark \kill
$q(X,Y) \ :- \ p(X,Y)$.
\end{tabbing}
\vspace{-0.4cm}
and for a set of dependencies $\Sigma$ with a single fd, $\Sigma = \{ \sigma: \ p(X,Y) \ \wedge \ p(X,Z) \rightarrow Y = Z \}$, an unchase step ``add to $Q$ a subgoal $p$ with a fresh variable for the second argument" can be applied infinitely many times. This query $Q'$ is a partial unchase result after two steps:
\vspace{-0.4cm}
\begin{tabbing}
This is my \= kill mark \kill
$q'(X,Y) \ :- \ p(X,Y), \ p(X,Z_1), \ p(X,Z_2)$.
\end{tabbing}
\vspace{-0.8cm}
\end{example}
\vspace{-0.4cm}

At the same time, we can guarantee unchase termination and ``good" properties of the {\tt cgalg} approach if we incorporate fds into unchase as follows: (1) An unchase step for fds is the same as a ``regular" chase step on fds, see Section~\ref{chase-prelim-section}. (2) A query is unchased in presence of fds combined with ids (tgds) by applying all the ids (tgds) before all the fds. The complexity of unchase under ids only is $m^3|\Sigma|$, where $m$ is the total number of subgoals in the query workload; the complexity of unchase under ids and fds is $m^4|\Sigma|$.

\vspace{-0.4cm}
\begin{proposition}
\label{tgd-fd-unchase-unique-proposition}
For a conjunctive query $Q$, for a set of dependencies $\Sigma$ that has fds either alone or in combination with ids or strongly acyclic tgds, and for any finite-size (either partial or terminal) chase result $Q'$ of $Q$ under $\Sigma$, $Q_{u,\Sigma}$ is equivalent to $Q'_{u,\Sigma}$ in the absence of dependencies.
\end{proposition}
\vspace{-0.4cm}
To prove this result, we apply and extend the id/fd separability result of~\cite{JohnsonK84} that says that $chase_{\Sigma[I+F]}(Q) \equiv chase_{\Sigma[I]}(chase_{\Sigma[F]}(Q))$ (for the notation, see Proposition~\ref{jk-proposition}).

This result is obtained using Proposition~\ref{tgd-fd-unchase-unique-proposition}:
\vspace{-0.4cm}
\begin{theorem}
For any two conjunctive queries $Q_1$ and $Q_2$ and for a set of dependencies $\Sigma$ that satisfies the conditions of Proposition~\ref{tgd-fd-unchase-unique-proposition}, $Q_1 \equiv_\Sigma Q_2$ if and only if $Q_{1,u,\Sigma}$ is equivalent to $Q_{2,u,\Sigma}$ in the absence of dependencies.
\end{theorem}
\vspace{-0.4cm}

To obtain beneficial reformulations for a problem input $\cal I$, we apply {\tt cgalg} on the terminal results of unchasing the workload queries in $\cal I$ under the set of dependencies in $\cal I$. 
\vspace{-0.4cm}
\begin{theorem}
\label{main-unchase-id-theorem}
{\tt cgalg$(\{ Q_{u,\Sigma} \},{\cal D},L)$} is sound for problem inputs $\cal I$ where the workload ${\cal Q} = \{ Q \}$ has conjunctive queries only and such that $\Sigma$ satisfies the conditions of Proposition~\ref{tgd-fd-unchase-unique-proposition}. The algorithm is complete for such problem inputs provided the queries $Q_{u,\Sigma}$ have no self-joins. 
\end{theorem}
\vspace{-0.4cm}

By definition of the unchase process, the complexity of {\tt cgalg} in this case is $O(2^m)$, where $m$ is the total number of subgoals in the workload queries in the problem input $\cal I$.

\vspace{-0.4cm}
\begin{theorem}
For problem inputs $\cal I$ that satisfy the conditions of Theorem~\ref{main-unchase-id-theorem}, the decision version of the problem of generating optimal reformulations is in NP, provided that the queries $unchase_\Sigma (Q)$ have no self-joins.
\end{theorem}
\vspace{-0.4cm}


It is remarkable that, given a problem input $\cal I$ and the rewritings produced by {\tt cgalg} on the terminal results of unchasing the queries in $\cal I$ using the dependencies in $\cal I$, to show the equivalence of the original workload queries to the rewritings, we {\it do not need} to unchase the expansions of the rewritings. (Note that one needs to apply {\it chase} to discover rewritings that are equivalent to queries under dependencies; see, e.g.,~\cite{DeutschLN05}. We can show that if we used the approach described in Section~\ref{dep-chase-section}, we {\it would} need to chase the rewriting expansions to show the equivalence of the rewritings to the original queries.)
\vspace{-0.4cm}
\begin{theorem}
For problem inputs $\cal I$ that satisfy the conditions of Theorem~\ref{main-unchase-id-theorem}, let $R$ be a reformulation of some query $Q$ in $\cal I$, such that $R$ is returned by {\tt cgalg$(\{ Q_{u,\Sigma} \},$} {\tt ${\cal D},L)$}. Suppose $R^{exp} \equiv Q_{u,\Sigma}$ in the absence of dependencies. Then $R^{exp}_{u,\Sigma} \equiv Q_{u,\Sigma}$ in the absence of dependencies.
\end{theorem}
\vspace{-0.4cm}

\vspace{-0.5cm}
\section{Conclusions; Future Work}
\vspace{-0.3cm}

\label{conclus-section}

We have presented complexity results and {\tt cgalg} algorithms for database reformulation in presence of dependencies. Our results apply to conjunctive queries and to the types of dependencies that include commonly used functional dependencies, inclusion dependencies, and foreign-key constraints. We argued that to generate beneficial reformulations, one can use the chase technique for incorporating dependencies into query definitions. At the same time, we showed that we can reduce the complexity of database reformulation {\it and} cover larger classes of dependencies by incorporating into the reformulation algorithm our unchase approach; the idea of unchase is to remove from a query all ``derived" subgoals that would be introduced by chase. 

The unchase/{\tt cgalg} approach can be extended to workloads of queries with self-joins, at the expense of an increase in runtime complexity (cf.~\cite{ChirkovaHS02,AfratiCGP05}). We are currently working on extending the approach to database reformulation for queries with aggregation. Another direction of our ongoing and future work is designing efficient algorithms for database reformulation for common classes of queries and dependencies. Besides database reformulation, the unchase approach can be used in answering queries using views, as it reduces the problem of checking query containment (equivalence) in presence of dependencies to the problem of containment (equivalence) checking in the absence of dependencies, {\it without} increasing query size. Note that unchase, unlike chase, can be used in presence of cyclic inclusion dependencies. Exploring unchase for answering queries using views is another direction of our future work.

\small
\bibliography{mrg}

\end{document}